\documentclass{ws-procs9x6}

\newcommand{\cc}[2]{c{#1\atopwithdelims[]#2}}

\newcommand{\nn}{\nonumber}

\begin{document}

\title{
\rightline{OUTP--03--32P}
\rightline{LPTENS-03/27}
\rightline{\tt hep-th/0311058}
\rightline{November 2003}
\rightline{}
Towards the classification of\\ $Z_2\times Z_2$ fermionic models
}



\author{A.E. Faraggi$^1$, C. Kounnas$^2$,
S.E.M. Nooij$^1$\footnote{Invited talk presented by {\sc s}ander {\sc n}ooij
at the Second International Conference on String Phenomenology, 
29/7-4/8 2003, Durham UK}
and
J. Rizos$^{2,3}$}

\address{$^1$Theoretical Physics, University of Oxford, Oxford, OX1 3NP, England}
\address{$^2$Laboratoire de Physique Theorique, Ecole Normale Sup\'erieure, 24 rue L'homond, F--75231 Paris Cedex 05, France}
\address{$^3$Department of Physics, University of Ioannina, 45110 Ioannina, Greece}


\maketitle

\abstracts{
We develop a formalism that allows a complete classification of
four-dimensional  $Z_2\times Z_2$ heterotic string models.
Three generation models in a sub--class of these compactifications
are related to the existence of three twisted sectors
in $Z_2\times Z_2$ orbifolds.
In the work discussed here we classify the sub--class of these
models that produce spinorial representations from 
all three twisted planes, and including symmetric shifts
on the internal lattice. We show that perturbative three
generation models are not obtained solely with symmetric shifts
on complex tori, but necessitate the action of an asymmetric shift.
In a subclass of these models we show that their chiral content
is predetermined by the choice of the $N=4$ lattice.
The implication of the results and possible geometrical interpretation
are briefly discussed.
}

\section{Introduction}
Some of the best studied heterotic string vacua up to date are the free fermionic
models \cite{FFF}.
They  describe a heterotic string compactified down to four dimensions.
The number of supersymmetries is reduced by orbifolding the internal space with a $Z_2 \times Z_2$ action.
A subclass of these models, the realistic free fermionic models, has been studied extensively in
the literature due to their interesting phenomenological features
\cite{Antoniadis:1989zy,Faraggi:1989ka,Antoniadis:1990hb,Faraggi:1991jr}.
Their chiral spectrum consists of three fermion generations with an $SO(10)$ embedding.
This $SO(10)$ is broken to one of its subgroups.
The effective low energy field theory can be directly calculated by
evaluating the associated string amplitudes.
Detailed investigation 
shows that they
may reproduce the basic features of the MSSM.
To achieve $SO(10)$ embedding a particular common set of
 basis vectors has been employed in all these models.
One to three additional Wilson-line like vectors were then utilized in order to achieve the $SO(10)$ breaking and the
family reduction.
In this work we examine the $SO(10)$ embedding in detail, starting from a more general basis 
and spans a wide class of free-fermionic models.
The basis vectors remain fixed and the various
models are generated by varying the projection coefficients, a set of discrete parameters associated with the
GSO projection. This permits a systematic analysis of the models and the production of algebraic formulas
for the main features of the models, like the number of generations.
A similar method has been utilized for the classification
of symmetric $Z_2\times Z_2$ Type II superstring vacua\cite{GKR}. Moreover, our formulation has the advantage
of separating the orbifold action from the family reduction mechanism giving new insight into various
puzzling issues, such as the chirality generation.
We find that in a big subclass of these models, all
major phenomenological features are determined at the $N=4$ level, offering a new understanding to the structure
of heterotic vacua that could be useful in the study of string dualities. The formalism developed  is also suitable
for a complete classification of this large class of models.
We report here the main lines of this formulation giving also some examples, a detail description and
model classification will be presented  elsewhere \cite{prep}.

In the free fermionic formulation the 4-dimensional heterotic string,  in the light-cone gauge, is described
by $20$ left moving  and $44$ right moving real fermions.
A large number of models can be constructed by choosing
different phases picked up by   fermions ($f_A, A=1,\dots,44$) when transported along
the torus non-contractible loops.
Each model corresponds to a particular choice of fermion phases consistent with modular invariance
that can be generated by a set of  basis vectors $v_i,i=1,\dots,n$
$$v_i=\left\{\alpha_i(f_1),\alpha_i(f_{2}),\alpha_i(f_{3}))\dots\right\}$$
describing the transformation  properties of each fermion
\begin{equation}
f_A\to -e^{i\pi\alpha_i(f_A)}\ f_A, \ , A=1,\dots,44
\end{equation}
The basis vectors span a space $\Xi$ which consists of $2^N$ sectors that give rise to the string spectrum. Each sector is given by
\begin{equation}
\xi = \sum N_i v_i,\ \  N_i =0,1
\end{equation}
The spectrum is truncated by a generalized GSO projection whose action on a string state  $|S>$ is
\begin{equation}\label{eq:gso}
e^{i\pi v_i\cdot F_S} |S> = \delta_{S}\ \cc{S}{v_i} |S>,
\end{equation}
where $F_S$ is the fermion number operator and $\delta_{S}=\pm1$ is the spacetime spin statistics index.
Different sets of projection coefficients $\cc{S}{v_i}=\pm1$ consistent with modular invariance give
rise to different models. Summarizing: a model can be defined uniquely by a set of basis vectors $v_i,i=1,\dots,n$
and a set of $2^{N(N-1)/2}$ independent projections coefficients $\cc{v_i}{v_j}, i>j$.

\section{The models in the free fermionic formulation}
\subsection{General setup}
The free fermions in the light-cone gauge in the usual notation are:
$\psi^\mu, \chi^i,y^i, \omega^i, i=1,\dots,6$ (left-movers) and  $\bar{y}^i,\bar{\omega}^i, i=1,\dots,6$,
$\psi^A, A=1,\dots,5$, $\bar{\eta}^B, B=1,2,3$, $\bar{\phi}^\alpha, \alpha=1,\ldots,8$ (right-movers).
The class of models we investigate, is generated by a set of 12 basis vectors
$$
B=\{v_1,v_2,\dots,v_{12}\},
$$
where
\begin{eqnarray}
v_1=1&=&\{\psi^\mu,\
\chi^{1,\dots,6},y^{1,\dots,6},\omega^{1,\dots,6}|\bar{y}^{1,\dots,6},\bar{\omega}^{1,\dots,6},\bar{\eta}^{1,2,3},
\bar{\psi}^{1,\dots,5},\bar{\phi}^{1,\dots,8}\},\nn\\
v_2=S&=&\{\psi^\mu,\chi^{1,\dots,6}\},\nn\\
v_{2+i}=e_i&=&\{y^{i},\omega^{i}|\bar{y}^i,\bar{\omega}^i\}, \ i=1,\dots,6,\nn\\
v_{9}=b_1&=&\{\chi^{34},\chi^{56},y^{34},y^{56}|\bar{y}^{34},\bar{y}^{56},\bar{\eta}^1,\bar{\psi}^{1,\dots,5}\},\label{basis}\\
v_{10}=b_2&=&\{\chi^{12},\chi^{56},y^{12},y^{56}|\bar{y}^{12},\bar{y}^{56},\bar{\eta}^2,\bar{\psi}^{1,\dots,5}\},\nn\\
v_{11}=z_1&=&\{\bar{\phi}^{1,\dots,4}\},\nn\\
v_{12}=z_2&=&\{\bar{\phi}^{5,\dots,8}\}.\nn
\end{eqnarray}
The vectors $1,S$ generate an
$N=4$ supersymmetric model. The vectors $e_i,i=1,\dots,6$ give rise
to all possible symmetric shifts of internal fermions
($y^i,\omega^i,\bar{y}^i,\bar{\omega}^i$) while $b_1$ and $b_2$
represent  the $Z_2\times Z_2$ orbifold twists. The remaining fermions not affected by the action
of the previous vectors are $\bar{\phi}^i,i=1,\dots,8$ which normally give rise to the hidden sector gauge group.
 The vectors $z_1,z_2$ divide these eight fermions into two sets of
four which in the $Z_2\times{Z_2}$ case is the maximum consistent partition\cite{FFF}.
This is the most general basis, with symmetric shifts for the internal fermions, that is compatible with a
Kac--Moody level one $SO(10)$ embedding.
 Without loss of generality we can set the associated projection coefficients
\begin{eqnarray}
\cc{1}{1}=\cc{1}{S}=\cc{S}{S}=\cc{S}{e_i}=\cc{S}{b_A}=-\cc{b_2}{S}=\cc{S}{z_n}=-1,
\label{asum}\end{eqnarray}
leaving  55 independent coefficients
\begin{eqnarray}
&&\cc{e_i}{e_j}, i\ge j, \ \ \cc{b_1}{b_2}, \ \ \cc{z_1}{z_2},\nn\\
&&\cc{e_i}{z_n}, \cc{e_i}{b_A},\cc{b_A}{z_n},
\ \ i,j=1,\dots6\,\ ,\  A,B,m,n=1,2\nn.
\end{eqnarray}
The remaining  projection coefficients are determined by modular invariance \cite{FFF}.
Each of the linearly independent coefficients can take two discrete values $\pm1$ and thus a simple counting
gives  $2^{55}$ (that is approximately $10^{16.6}$) distinct models in the class under consideration.

\subsection{The spectrum}
In a generic model described above the gauge group has the form
$$
SO(10)\times{U(1)}^3\times{SO(8)}^2
$$
Depending on the  choices of the projection coefficients, extra gauge bosons arise from
$
x=1+S+\sum_{i=1}^{6}e_i+z_1+z_2=\{{\bar{\eta}^{123},\bar{\psi}^{12345}}\}
$
resulting to enhancement $SO(10)\times{U(1)}\to E_6$. Additional gauge bosons can
arise  from the sectors
 $z_1,z_2$ and $z_1+z_2$ and  enhance
 ${SO(8)}^2\to SO(16)$ or ${SO(8)}^2\to E_8$. For particular choices of the projection coefficients
other gauge groups can be obtained \cite{prep}.

The untwisted sector matter is common to all models (putting aside gauge group enhancements)
and consists of six vectors of $SO(10)$ and  12 non-Abelian gauge group singlets.
Chiral twisted matter arises from the following 48 sectors (16 per orbifold plane)
\begin{eqnarray}
B_{\ell_3^1\ell_4^1\ell_5^1\ell_6^1}^1&=&S+b_1+\ell_3^1 e_3+\ell_4^1 e_4 +
\ell_5^1 e_5 + \ell_6^1 e_6 \nn\\
B_{\ell_1^2\ell_2^2\ell_5^2\ell_6^2}^2&=&S+b_2+\ell_1^2 e_1+\ell_2^2 e_2 +
\ell_5^2 e_5 + \ell_6^2 e_6 \label{ss}\\
B_{\ell_1^3\ell_2^3\ell_3^3\ell_4^3}^3&=&
S+b_3+ \ell_1^3 e_1+\ell_2^3 e_2 +\ell_3^3 e_3+ \ell_4^3 e_4\nn
\end{eqnarray}
where $\ell_i^j=0,1$ and $b_3=1+S+b_1+b_2+\sum_{i=1}^6 e_i+\sum_{n=1}^2 z_n$.
These states are  spinorials of $SO(10)$ and one can obtain at maximum one spinorial ($\bf 16$ or
$\bf {\overline{{16}}}$) per sector and thus totally 48 spinorials.
Extra non chiral matter i.e. vectors of $SO(10)$ as well as singlets arise from the sectors
$S+b_i+b_j+e_m+e_n$.

In our formulation we have separated the spinorials, that is we have separated the 48 fixed points
of the $Z_2\times{Z_2}$ orbifold. This separation allows us to examine the GSO action,
depending on the projection coefficients, on each spinorial separately. The choice of these coefficients
determines which  spinorials are  projected out, as well as
the chirality of the surviving states.

One of the  advantages of our formulation is that it allows to extract generic  formulas regarding the
number and the chirality of each spinorial. This is important because it allows an algebraic
treatment of the entire class of models without deriving each model explicitly. The number of
surviving spinorials per sector (\ref{ss}) is given by
\begin{eqnarray}
P_{\ell_3^1\ell_4^1\ell_5^1\ell_6^1}^{(1)}&=&
\frac{1}{16}\,\prod_{i=1,2}\left(1-\cc{e_i}{B_{\ell_3^1\ell_4^1\ell_5^1\ell_6^1}^{(1)}}\right)\,
\prod_{m=1,2}\left(1-\cc{z_i}{B_{\ell_3^1\ell_4^1\ell_5^1\ell_6^1}^{(1)}}\right)\,\label{pa}\\
P_{\ell_1^2\ell_2^2\ell_5^2\ell_6^2}^{(2)}&=&
\frac{1}{16}\,\prod_{i=3,4}\left(1-\cc{e_i}{B_{\ell_1^2\ell_2^2\ell_5^2\ell_6^2}^{(2)}}\right)\,
\prod_{m=1,2}\left(1-\cc{z_m}{B_{\ell_1^2\ell_2^2\ell_5^2\ell_6^2}^{(2)}}\right)\,\label{pb}
\\
P_{\ell_1^3\ell_2^3\ell_3^3\ell_4^3}^{(3)}&=&
\frac{1}{16}\,\prod_{i=5,6}\left(1-\cc{e_i}{B_{\ell_1^3\ell_2^3\ell_3^3\ell_4^3}^{(3)}}\right)\,
\prod_{m=1,2}\,\left(1-\cc{z_m}{B_{\ell_1^3\ell_2^3\ell_3^3\ell_4^3}^{(3)}}\right)\,\label{pc}
\end{eqnarray}
and thus the total number of spinorial per model is the sum of the above.
The chirality of the surviving spinorials is given by
\begin{equation}
X_{\ell_3^1\ell_4^1\ell_5^1\ell_6^1}^{(1)}=\cc{b_2+(1-\ell_5^1) e_5+(1-\ell_6^1) e_6}{B_{\ell_3^1\ell_4^1\ell_5^1\ell_6^1}^{(1)}}\label{ca}
\end{equation}
\begin{equation}
X_{\ell_1^2\ell_2^2\ell_5^2\ell_6^2}^{(2)}=\cc{b_1+(1-\ell_5^2) e_5+(1-\ell_6^2) e_6}{B_{\ell_1^2\ell_2^2\ell_5^2\ell_6^2}^{(2)}}\label{cb}
\end{equation}
\begin{equation}
X_{\ell_1^3\ell_2^3\ell_3^3\ell_4^3}^{(3)}=\cc{b_1+(1-\ell_3^3) e_3+(1-\ell_4^3) e_4}{B_{\ell_1^3\ell_2^3\ell_3^3\ell_4^3}^{(3)}}\label{cc}
\end{equation}
The net number of families is then
\begin{equation}
N=-\sum_{I=1}^3\sum_{p,q,r,s=0}^1X^{(I)}_{pqrs} P_{pqrs}^{(I)}
\end{equation}
Similar formulas can be easily derived for the number of vectorials and the number of singlets and can be extended
to the $U(1)$ charges.

Formulas (\ref{pa})-(\ref{pc}) allow us to identify the mechanism of spinorial reduction, or in other words
the fixed point reduction, in the fermionic language. For a particular sector ($B_{pqrs}$) of the orbifold plane $i$
there exist two shift vectors ($e_{2i-1}, e_{2i}$)
and the two zeta vectors ($z_1,z_2$) that have no common elements with $B_{pqrs}$. Setting the relative projection coefficients (\ref{pc})
to $-1$, each of the above four vectors acts as a projector that cuts the number of fixed points in the
associated sector by a factor of two. Since four such projectors are available for each sector the number of fixed points
can be reduced from 16 to one per plane.

\section{Models}
\subsection{The $Z_2\times{Z_2}$ symmetric orbifold}
The  configuration described above reproduces in the free fermionic language the $Z_2\times{Z_2}$ symmetric
orbifold model for the following choice of projection coefficients $\cc{v_i}{v_j}=\exp[i\pi(v_i|v_j)]$
{\footnotesize$$
(v_i|v_j)\ \ =\ \
\bordermatrix{&1   &  S  &  e_1  &   e_2   &  e_3   &  e_4  &   e_5  &   e_6   &  b_1   &  b_2   &  z_1  &   z_2\cr
   1  &  1   &   1  &    1  &    1  &    1  &    1  &    1   &   1   &   1  &    1  &    1  &   1 \cr
   S  & 1 &  1 &  1 &  1 &  1 &  1 &  1 &  1 &  1 &  1 &  1 & 1\cr
  e_1  & 1 &  1 &  0 &  0 &  0 &  0 &  0 &  0 &  0 &  0 &  0 & 0\cr
  e_2  & 1 &  1 &  0 &  0 &  0 &  0 &  0 &  0 &  0 &  0 &  0 & 0\cr
  e_3  & 1 &  1 &  0 &  0 &  0 &  0 &  0 &  0 &  0 &  0 &  0 & 0\cr
  e_4  & 1 &  1 &  0 &  0 &  0 &  0 &  0 &  0 &  0 &  0 &  0 & 0\cr
  e_5  & 1 &  1 &  0 &  0 &  0 &  0 &  0 &  0 &  0 &  0 &  0 & 0\cr
  e_6  & 1 &  1 &  0 &  0 &  0 &  0 &  0 &  0 &  0 &  0 &  0 & 0\cr
  b_1  & 1 &  0 &  0 &  0 &  0 &  0 &  0 &  0 &  1 &  0 &  0 & 0\cr
  b_2  & 1 &  0 &  0 &  0 &  0 &  0 &  0 &  0 &  0 &  1 &  0 & 0\cr
  z_1  & 1 &  1 &  0 &  0 &  0 &  0 &  0 &  0 &  0 &  0 &  1 & 1\cr
  z_2  & 1 &  1 &  0 &  0 &  0 &  0 &  0 &  0 &  0 &  0 &  1 & 1\cr
  }
$$}
The above choice wipes out the action of all projectors and thus the number of fixed points takes its maximum
value which is 48. The gauge group enhances to $E_6\times{U(1)}^2\times{E_8}$ and the spinorials of $SO(10)$
combine with vectorials and singlets to produce 48+3=51 families ($\bf 27$) and 3 anti-families ($\bf \overline{27}$) of $E_6$.

\subsection{A three family $SO(10)$ model\label{subs}}
We can obtain a three family $SO(10)$ model by choosing the following set of projection coefficients
{\footnotesize$$
(v_i|v_j)\ \ =\ \
\bordermatrix{
       &1  &S   & e_1& e_2& e_3& e_4&e_5 & e_6& b_1& b_2&z_1 &z_2\cr
   1   &1  &  1 &  1 &  1 &  1 &  1 &  1 &  1 &  1 &  1 &  1 & 1 \cr
   S   & 1 &  1 &  1 &  1 &  1 &  1 &  1 &  1 &  1 &  1 &  1 & 1\cr
  e_1  & 1 &  1 &  0 &  0 &  0 &  1 &  1 &  0 &  0 &  0 &  0 & 0\cr
  e_2  & 1 &  1 &  0 &  0 &  1 &  0 &  1 &  1 &  0 &  0 &  0 & 0\cr
  e_3  & 1 &  1 &  0 &  1 &  0 &  0 &  0 &  0 &  0 &  0 &  0 & 1\cr
  e_4  & 1 &  1 &  1 &  0 &  0 &  0 &  0 &  0 &  0 &  0 &  1 & 0\cr
  e_5  & 1 &  1 &  1 &  1 &  0 &  0 &  0 &  1 &  0 &  0 &  0 & 1\cr
  e_6  & 1 &  1 &  0 &  1 &  0 &  0 &  1 &  0 &  0 &  0 &  1 & 0\cr
  b_1  & 1 &  0 &  0 &  0 &  0 &  0 &  0 &  0 &  1 &  0 &  0 & 0\cr
  b_2  & 1 &  0 &  0 &  0 &  0 &  0 &  0 &  0 &  0 &  1 &  0 & 0\cr
  z_1  & 1 &  1 &  0 &  0 &  0 &  1 &  0 &  1 &  0 &  0 &  1 & 1\cr
  z_2  & 1 &  1 &  0 &  0 &  1 &  0 &  1 &  0 &  0 &  0 &  1 & 1\cr
  }
$$}
The gauge group is here $SO(10)\times {U(1)}^3\times {SO(8)}^2$ and the observable sector consists
of three $SO(10)$ spinorials (${\bf 16}$) (one from each sector) and  nine $SO(10)$ vectors (${\bf 10}$).
The hidden sector consists of  nine octoplets under each $SO(8)$. In addition there exist 40 non-abelian
gauge group singlets. The model could be phenomenologically acceptable provided one finds a way
of breaking $SO(10)$. Since it is not possible to generate the $SO(10)$ adjoint (not in Kac-Moody level one),
we need to realize the breaking by an additional Wilson-line like vector. However,  a detailed
investigation of acceptable basis vectors, shows that the $SO(10)$ breaking is accompanied by
truncation of the fermion families. Thus this kind of perturbative $SO(10)$ breaking is not compatible
with the presence of three generations. It would be interesting to utilize string dualities in order to study
the non-perturbative aspects of such models.

\section{Connection with realistic models}
All realistic heterotic free fermionic models constructed so far are builded using on a common set of 8 basis vectors
labeled $\{1, S, b_1, b_2, b_3, b_4, b_5,\alpha\}$.
The first four vectors ($\{1, S, b_1, b_2\}$) coincide with  $\{v_1,\dots v_4\}$ of our basis (\ref{basis})
 while $b_3$ is a linear combination of the total sum
\begin{equation}\label{b3}
b_3=\sum_{i=1}^{12} v_i=1+S+b_1+b_2+\sum_{i=1}^6 e_i+\sum_{n=1}^2 z_n
\end{equation}
The vectors $b_4$ and $b_5$ can be written as linear combinations of $e_i$'s using
\begin{equation}
b_1+b_4=v_6+v_7=e_4+e_5\ ,\ b_2+b_5=v_3+v_8=e_1+e_6\
\end{equation}
The vector
\begin{equation}
\alpha=1+S+\sum_{i=1}^{6}e_i+z_1
\end{equation}
corresponds to $2\alpha$ of \cite{Antoniadis:1989zy},  
$\alpha$ of \cite{Antoniadis:1990hb} and
$2\gamma$ of \cite{Faraggi:1991jr}. 
Thus our formulation reproduces all 8 common basis vectors
by exchanging, for example, $z_1,z_2$ of our formalism for $b_3$ and 
$\alpha$. However, four
additional vectors are present in our basis 
(four linear combinations of the $e_i$'s). These shifts
simply play the role of projectors, depending on the value of the 
associated projection coefficients,
and do not give rise of additional massless states. Inactive shift vectors,
can be removed from
the basis without modifying the model's spectrum and this explains the
existence of $SO(10)$ embeddings  in
e.g \cite{Antoniadis:1989zy,Antoniadis:1990hb,Faraggi:1991jr}
with fewer shift vectors.

 Some extra vectors (one to three depending on
 the model)  are usually employed, in the case of realistic models, in order to
 achieve the final reduction of the number of families and the gauge group to either MSSM or an $SO(10)$ subgroup
containing it. These vectors are not relevant to a $SO(10)$ embedding but to a $SO(10)$ breaking, so we will
not consider  them here since the problem of the $SO(10)$ breaking will be discussed in a future publication.

\section{$N=4$ liftable models}
In the models considered above we have managed to separate the  orbifold twist action (represented here by $b_1$, $b_2$) from
the shifts (represented by $e_i$) and the Wilson lines $z_1,z_2$. However, these actions are further correlated
through the GSO projections.  Nevertheless, we  remark that these action are completely decoupled in the case
\begin{equation}
\cc{b_i}{z_m}=\cc{b_i}{e_i}=+1
\end{equation}
In this subclass of models  the orbifold action is restricted to the reduction of supersymmetry
and the gauge group. The important phenomenological properties of the model, as the number of generations, are
predetermined at the $N=4$ level. A typical example of such a model is the three generation $SO(10)\times {U(1)}^3\times {SO(8)}^2$ presented
in \ref{subs}.
This class of models consists in principle of $2^{39}$ models. Taking into account some symmetries among the
coefficients (for example the operation $\cc{b_1}{b_2}\to-\cc{b_1}{b_2}$ simply changes the overall chirality)
the number of models can be further reduced and can be classified in detail using our analytical formulas and a computer
program. Detailed tables presenting all different models in this subclass together with their generation number
have been derived and will be reported in a future publication \cite{prep}.
This category of models is of particular interest since they admit a geometric interpretation.

\section{Conclusions}
We have developed a formalism that describes the whole class of fermionic heterotic vacua with a $SO(10)$ embedding
in terms of a fixed set of basis vectors and a set of discrete binary parameters associated with GSO projections.
Our formalism permits a systematic classification of these models and detailed investigation of their
main phenomenological properties. A subclass of these vacua turns out to be of special interest as their
phenomenology is inherited from the  $N=4$ theory obtained when removing the orbifold twist. We have presented
some typical examples of models derived in our formulation and analytic formulas for the generation number.
We have also identified in technical level the mechanism of generation reduction necessary to obtain the
phenomenologically acceptable models.
Preliminary results show that, in the framework of  $Z_2\times Z_2$ Calabi--Yau
compactification, the net number of generation  can never be equal to
three. This implies the necessity of non zero torsion in CY
compactifications in order to obtain semi-realistic three
generation models. In a future publication we
will study in detail this class of vacua and consider also the problem of
$SO(10)$-breaking. The necessity to incorporate an asymmetric shift
in the reduction to three generations, has profound implications for the
issues of moduli stabilization and vacuum selection. The reason
being that it can only be implemented at enhanced symmetry
points in the moduli space. In this context we envision that the
self--dual point under T--duality plays a special role. In the 
context of nonperturbative dualities the dilaton and
moduli are interchanged, with potentially important
implications for the problem of dilaton stabilization.
We note that the low energy phenomenological data
may point in the direction of esoteric compactifications that
would have otherwise been overlooked.
We will report on these aspects in future publications. 

\section*{Acknowledgements}
This  work was supported in part by the Pieter Langerhuizen
Lambertuszoon Fund of the Royal Holland Society of Sciences and Humanities,
the Noorthey Academy and the VSB foundation(SN); by the  European Union
under the
contracts HPRN-CT-2000-00122, HPRN-CT-2000-00131,
HPRN-CT-2000-00148, HPRN-CT-2000-00152 and HPMF-CT-2002-01898; 
by the PPARC and by the Royal Society.

\end{document}